\documentstyle[twoside]{article}
\newcount\Mac  \Mac=0  
\newcommand{\ifMac}[2]{\ifnum\Mac=1 #1 \else #2 \fi}
\def\Red  {}
\def\Black{}
\def\Blue {}
\newcommand{\GeV}{\,{\rm GeV}}

\newcommand{\NP}{Nucl. Phys.}
\newcommand{\PRL}{Phys. Rev. Lett.}
\newcommand{\PL}{Phys. Lett.}
\newcommand{\PR}{Phys. Rev.}

\def\circa#1{\,\raise.3ex\hbox{$#1$\kern-.75em\lower1ex\hbox{$\sim$}}\,}
\makeatletter
\def\art{\@ifnextchar[{\eart}{\oart}}
\def\eart[#1]#2#3#4#5#6{{\rm #2}, {\em #3 \bf #4} {\rm (#6) #5}}
\def\hepart[#1]#2{{\rm #2, \em#1}}
\newcommand{\oart}[5]{{\rm #1}, {\em #2 \bf #3} {\rm (#5) #4}}

\newcounter{alphaequation}[equation]
\def\thealphaequation{\theequation\hbox to
0.6em{\hfil\alph{alphaequation}\hfil}}
\def\eqnsystem#1{
\def\@eqnnum{{\rm (\thealphaequation)}}
\def\@@eqncr{\let\@tempa\relax \ifcase\@eqcnt \def\@tempa{& & &} \or
  \def\@tempa{& &}\or \def\@tempa{&}\fi\@tempa
  \if@eqnsw\@eqnnum\refstepcounter{alphaequation}\fi
\global\@eqnswtrue\global\@eqcnt=0\cr}
\refstepcounter{equation} \let\@currentlabel\theequation \def\@tempb{#1}
\ifx\@tempb\empty\else\label{#1}\fi
\refstepcounter{alphaequation}
\let\@currentlabel\thealphaequation
\global\@eqnswtrue\global\@eqcnt=0 \tabskip\@centering\let\\=\@eqncr
$$\halign to \displaywidth\bgroup \@eqnsel\hskip\@centering
$\displaystyle\tabskip\z@{##}$&\global\@eqcnt\@ne
\hskip2\arraycolsep\hfil${##}$\hfil& \global\@eqcnt\tw@\hskip2\arraycolsep
$\displaystyle\tabskip\z@{##}$\hfil
\tabskip\@centering&\llap{##}\tabskip\z@\cr}
\def\endeqnsystem{\@@eqncr\egroup$$\global\@ignoretrue} \makeatother

\begin{document}
\twocolumn[
\centerline{20 Jan.\ 1998 \hfill    IFUP--TH/4--98}
\centerline{hep-ph/9801353 \hfill SNS-PH/1998-2} \vspace{1cm}
\centerline{\LARGE\bf\Red About the fine-tuning price of LEP}

\bigskip\bigskip\Black
\centerline{\large\bf Riccardo Barbieri {\rm and} Alessandro Strumia} \vspace{0.3cm}

\centerline{\em Dipartimento di Fisica, Universit\`a di Pisa and}
\centerline{\em INFN, sezione di Pisa,  I-56126 Pisa, Italia}\vspace{0.3cm}

\bigskip\bigskip\Blue

\centerline{\large\bf Abstract}
\begin{quote}\large\indent
Following Chankowski, Ellis and Pokorski we quantify the amount of fine-tuning
of input parameters of the
Minimal Supersymmetric Standard Model
that is needed to respect the lower limits on sparticle and Higgs masses
imposed by negative searches so far, direct or indirect.
By including the one loop radiative corrections to the
effective potential,
the amount of fine-tuning is reduced with respect to the results of CEP by
a factor of $2\div 5$, strongly increasing as $\tan\beta$ approaches 1.
A further reduction factor may come from a more appropriate, less restrictive,
definition of the fine-tuning parameter itself.
\end{quote}\Black
\vspace{1cm}]

\noindent
Extensive searches of supersymmetric signals, done mostly, but not only,
at LEP, explore in a significant way the parameter space of the Minimal
Supersymmetric Standard Model (MSSM).
Although variations on the MSSM are certainly possible, the MSSM is, in some
of its aspects, representative enough to make these searches relevant in absolute terms,
hence the importance of assessing their significance in a quantitative way.
In essence, how much should one worry about the fact that no positive result has been
found so far?
Or, even more importantly, how critical are, to test the MSSM, the searches that
will be performed in the nearest future?

A first important attempt to answer these questions has been recently made by
Chankowski, Ellis and Pokorski~\cite{CEP} (CEP), who study the minimal
`amount of fine-tuning'~\cite{FT} $\Delta$ needed to live outside of
the MSSM parameter space excluded by the most significant negative searches of
supersymmetry done so far.
The data taken into account by CEP include:
\begin{enumerate}
\item  the latest set of precision electroweak tests~\cite{EWtests};
\item the latest measurement of the branching ratio
${\rm B.R.}(B\to X_s\gamma)$~\cite{bsgExp} and,
most importantly
\item the lower limits on sparticle and Higgs boson masses from LEP2~\cite{LEP2}.
\end{enumerate}
A way to summarise the results of CEP is the following:
\begin{itemize}
\item An amount of fine tuning $\Delta$ greater than about 20 is required for any value of
$\tan\beta$;
\item A scalar Higgs lighter than $90\GeV$, that will be searched for at LEP in the near future,
requires $\Delta\circa{>}60$;
\item A value of $\tan\beta$ lower than 2, a range 
suggested by an infra-red fixed-point analysis, requires $\Delta$ greater than about $100$.
\end{itemize}
Since $\Delta$ is supposed to measure, although in a rough way, the inverse
probability of an unnatural cancellation to occur
in the expression that relates the $Z$-boson mass to the various MSSM parameters,
the results of CEP are striking enough to suggest that we take a closer look at them,
which is what we do in this letter.
This is without underestimating the difficulty of giving an unambiguous quantitative
meaning to the `naturalness bounds' on the sparticle masses,
as correctly emphasised by CEP themselves.

Our results are summarised in fig.~1, which gives the prize of the fine tuning
$\Delta$ as a scatter plot in the MSSM parameters space still consistent
with the data mentioned above and considered by CEP\footnote{The density of points in this plot
does not have any particular meaning. What counts is rather the enveloping curve at the lower bound
of the populated region.}.
As seen from fig.~1, we have a lower limit on $\Delta$ which is about a factor
of $5\div10$ weaker than that obtained by CEP,
a relative factor actually increasing as $\tan\beta$ gets close to~1.
This is mostly due to the inclusion of the full one-loop corrections to the
scalar potential.
The relative factor between CEP and us also includes an overall factor of about~2 for
every $\tan\beta$, coming from what is probably a more adequate definition of the fine tuning
parameter itself than the one adopted by CEP (see below).

\begin{figure}[t]\setlength{\unitlength}{1cm}
\begin{center}\begin{picture}(8.5,8)
\ifMac{\put(-0.5,0){\special{picture FTb}}}
{\put(-0.5,0){\includegraphics{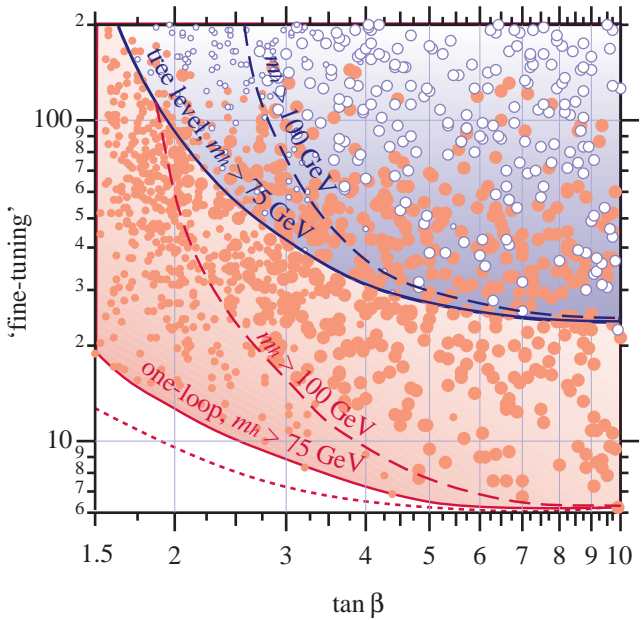}}}
\end{picture}
\caption{\em Scatter plot of the fine-tuning as function of $\tan\beta$.
In the empty $\Blue\circ\Black$ (filled $\Red\bullet\Black$) points
the FT is computed as in CEP (as here).
Small points have $m_h> 75 \GeV$, bigger points have $m_h>100\GeV$.
The lower dotted line is explained in the text.}
\end{center}
\end{figure}

\begin{figure}[t]\setlength{\unitlength}{1cm}
\begin{center}\begin{picture}(8.5,8)
\ifMac{\put(-0.5,0){\special{picture red}}}
{\put(-0.5,0){\includegraphics{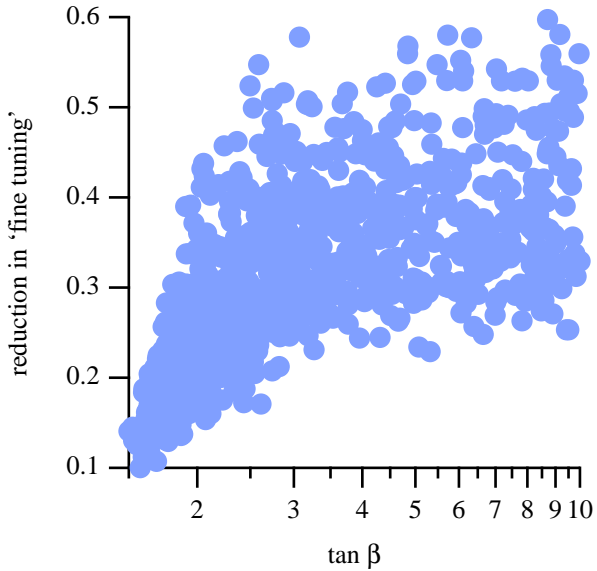}}}
\end{picture}
\caption{\em Reduction in fine-tuning due to one-loop effects.
The parameters of the tree level potential have been renormalized at
$Q=175\GeV$.\label{fig:red}}
\end{center}
\end{figure}

\begin{figure}[t]\setlength{\unitlength}{1cm}
\begin{center}\begin{picture}(8.5,8)
\ifMac{\put(-0.5,0){\special{picture FTh}}}
{\put(-0.5,0){\includegraphics{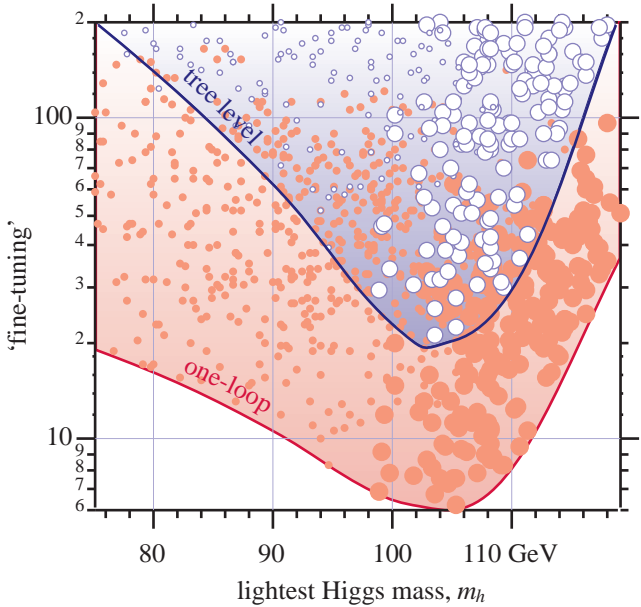}}}
\end{picture}
\caption{\em Scatter plot of the fine-tuning as function of $m_h$.
In the empty $\Blue\circ\Black$ (filled $\Red\bullet\Black$) points
the FT is computed as in CEP (as here).
Small points have $\tan\beta<4$, bigger points have $\tan\beta>4$.}
\end{center}
\end{figure}

The inclusion of the full one-loop corrections to the scalar potential~\cite{V1loop}
is known to be an
essential ingredient to make the result not too sensitive to the choice of
the renormalization scale of the various low energy parameters~\cite{V1loopQ}.
In turn, this has the effect of reducing the fine-tuning parameter especially for
values of $\tan\beta$ near to one (where the tree level supersymmetric potential
has a flat direction)
and for high values of $\Delta$
(because there is a large spread in the superpartner spectrum).
Since this is the main source of the difference between CEP and us,
this relative reduction effect is shown in fig.~\ref{fig:red},
also in the form of a scatter plot, for a fixed definition of $\Delta$,
no matter which.

\medskip

The strong increase of the {\em tree-level\/} fine-tuning at small
values of $\tan\beta$ is not due to the fact that
the experimental constraints are stronger at small $\tan\beta$,
but to the particular nature of the {\em tree level\/} MSSM potential:
its super-gauge part vanishes at $\tan\beta=1$.
For this reason, {\em in tree-level approximation}
the second derivative of the potential is small
at $\tan\beta\sim1$,
$V''\sim M_Z^2\cos^2 2\beta\sim m_h^2$,
and a variation of the MSSM parameters
induces a very large variation in $M_Z^2$, dominantly through
the induced variation of $\tan\beta$.
To illustrate this point,
we have also computed the logarithmic derivative that defines the fine tuning
{\em at fixed values of $\tan\beta$}\footnote{This different choice of
the basic `input' parameters can also be a correct one.
For example the pseudo-Goldstone mechanism for a light Higgs doublet~\cite{PG}
links the tree level
parameters of the Higgs potential so that $\tan\beta=1$.}.
The minimal experimentally allowed value of this biased fine tuning,
represented by the dotted line in fig.~1,
does not strongly increase at small $\tan\beta$, and
is comparable to the one-loop unbiased value (continuous line).
The reduction in fine-tuning due to one-loop effects is larger if the sign
of $\mu$ is positive and is roughly given by
$\Delta_{\rm loop}/\Delta_{\rm tree}=m_{h\rm tree}^2/m_{h\rm loop}^2$.

\medskip

The definition of $\Delta$ in terms of the logarithmic sensitivities of $M_Z$
with respect to variations of a set of input parameters $\{\wp\}$
($\mu$-term,
soft terms, gauge and Yukawa couplings) chosen to be the ``fundamental'' ones
has been criticized~\cite{FTcritica} as being too restrictive or unadequate at all.
Different definitions give bounds weaker by a factor that can go up to $3\div 4$.
Partly, this is where the ambiguity of the quantitative concept comes in.
We stick to a definition~\cite{GMFT} that avoids these criticisms by replacing the
logarithmic derivatives present in the original definition
$$d=\left|\frac{\mu}{M_Z^2}\frac{d M_Z^2}{d \mu}\right|$$
with a particular finite difference
$$\Delta=\left|\frac{\mu^2}{M_Z^2}\frac{\Delta M_Z^2}{\Delta \mu^2}\right|$$
explicitly defined in~\cite{GMFT}.
In practice $2\Delta$ reduces to $d$ when $\Delta\gg 1$
and is up to 30\% lower than it in all cases of interest.
This definition more directly weights as unnatural the possibility of cancellations
between different contributions to the expression of the $Z$-boson mass.
To make a long story short, we say that there is an unnaturally small probability
$p\approx \Delta^{-1}$ that every single `contribution' to
$M_Z^2$ be $\Delta$ time bigger than their sum, $M_Z^2$.
This explains a factor of~2 between us and CEP.
The fine-tuning with respect to the `$\mu$-term' parameter
is the one that gives the strongest constraint.

\bigskip

As pointed out by CEP, a key search to be performed
in the nearest future is the one of the lightest MSSM Higgs boson at LEP2.
The impact of this search is illustrated in fig.~3 for every value of
$\tan\beta$ less than~10.
The analysis for large values of $\tan\beta$ is beyond the scope of this letter.
The low $\Delta$-values for $m_h\circa{>}90\GeV$ in fig.~3 correspond to moderately large
values of $\tan\beta$ (but still below 10).
To explore the $h$-mass region up to $100\GeV$ is known to be a very significant study
of the low $\tan\beta$ region.
This is illustrated by comparing the dashed lines with the
continuous ones in fig.~1,
where the only difference is the constraint
that $m_h> 100\GeV$ (dashed lines)
rather than the present limit, $m_h> 75\GeV$ (continuous lines).
The values of $m_h$ have been computed using the approximation presented in~\cite{mhApprox}.

\medskip

As stressed in~\cite{CEP} the fine-tuning bounds are not much different
in non minimal supergravity models~\cite{DG}.
In gauge-mediation models
the bound on the right-handed slepton masses becomes 
the relatively most relevant constraint.
Its main effect is that, in minimal gauge mediation,
$\Delta<10$ can only be achieved if the messenger scale $M_M$
is sufficiently high, $M_M\circa{>}10^{12}\GeV$, or
extremely low~\cite{GMFT}.

\bigskip

In conclusion we confirm the significance of
the negative searches performed so far to look for a supersymmetric
particle spectrum in a direct or an indirect way.
In relative terms, adopting a quantitative measure of the naturalness criterium
to assess their significance, the searches for
charginos and neutralinos at LEP2 have played the most significant role so far.
Nevertheless, regions of the MSSM parameter space with a relatively high
``naturalness probability'' still exist.
The need to explore them confirms the importance of looking at LEP2 for the
highest possible Higgs masses.

\end{document}
\\
Title: About the fine-tuning price of LEP
Authors: Riccardo Barbieri and Alessandro Strumia
Comments: 3 pages.
Report-no: SNS-PH/1998-2
\\
Following Chankowski, Ellis and Pokorski we quantify the amount of fine-tuning
of input parameters of the
Minimal Supersymmetric Standard Model
that is needed to respect the lower limits on sparticle and Higgs masses
imposed by negative searches so far, direct or indirect.
By including the one loop radiative corrections to the
effective potential,
the amount of fine-tuning is reduced with respect to the results of CEP by
a factor of $2\div 5$, strongly increasing as $\tan\beta$ approaches 1.
A further reduction factor may come from a more appropriate, less restrictive,
definition of the fine-tuning parameter itself.
\\